\documentclass[twocolumn,showpacs,amsmath,amssymb,pra]{revtex4}
\usepackage{textcomp}
\usepackage{amsmath}
\usepackage{amssymb}
\usepackage{graphicx}
\usepackage{epstopdf}
\usepackage{bbm}
\makeatletter

\usepackage{textcomp}
\usepackage{epstopdf}
\usepackage{bbm}
\usepackage[toc,page,title,titletoc,header]{appendix}
\usepackage{dcolumn}\usepackage{bm}
\newcommand{\be}{\begin{equation}}
\newcommand{\ee}{\end{equation}}
\newcommand{\bey}{\begin{eqnarray}}
\newcommand{\eey}{\end{eqnarray}}
\newcommand{\bw}{\begin{widetext}}
\newcommand{\ew}{\end{widetext}}

\newcommand{\ov}{\overline}

\newcommand{\ra}{\rangle}
\newcommand{\la}{\langle}

\newcommand{\ba}{\begin{array}}
\newcommand{\ea}{\end{array}}
\newcommand{\bi}{\begin{itemize}}
\newcommand{\ei}{\end{itemize}}
\newcommand{\bem}{\begin{enumerate}}
\newcommand{\eem}{\end{enumerate}}

\makeatother

\begin{document}

\title{Internal temperature of quantum chaotic systems at the nanoscale
 and its detection by a microscopic thermometer}

\author{Jiaozi Wang and Wen-ge Wang \footnote{ Email address: wgwang@ustc.edu.cn}}

\affiliation{Department of Modern Physics, University of Science and Technology
of China, Hefei, 230026, China}

\date{\today}
\begin{abstract}
 The extent to which a temperature can be appropriately assigned to a small quantum system, as an internal
 property but not as a property of any large environment, is still an open problem. In this
 paper, a method is proposed for solving this problem, by which
 a studied system is coupled to a two-level system (probe) as a microscopic thermometer.
 For small quantum chaotic systems, we show that a temperature can be determined, the value of which
 is sensitive to neither the form, location, and strength of
 the probe-system coupling, nor the Hamiltonian and initial state of the probe.
 This temperature turns out to have the form of Boltzmann temperature.
\end{abstract}



\maketitle

\section{Introduction}
 Thermal and statistical properties of small quantum systems have been receiving
 lots of attention in recent years, both theoretical and experimental
 \cite{Heatengine_thr1,Heatengine_thr2,Heatengine_thr3,Heatengine_exp1,gogolin,Eisert14,Hartmann04,
 Rapp10,Braun13,Ne_Temp_thr1,Ne_Temp_thr2,Ne_Temp_thr3,Ne_Temp_thr4,Ne_Temp_exp1,Rossini16}.
 A key concept in this field, namely, temperature for such systems has not been fully understood, yet
 \cite{Eisert14,Hartmann04,Rossini16}.
 In particular, for a small quantum system, which possesses nonweak interactions among its components
 and is approximately isolated from its environment, the extent to which a temperature can be assigned to it,
 as an internal property but not as a property of environment, is still an open problem.
 To solve this problem is a challenge to both theoretical and experimental physics.

 On one hand, in the statical mechanics, temperature can be defined in several ways, which are equivalent
 in the thermodynamic limit,  e.g., that by Boltzmann's entropy \cite{Landau}
 and that by Gibbs' entropy \cite{Dunkel14}.
 But,  there is by far no unique way for extrapolation to small quantum systems
 \cite{Dunkel14,Izrailev-old,Izrailev-new,Casati-th1,Casati-th2,ethbath,Genway12}.
 Different understandings of this concept may lead to diverse predictions; for example,
 related to the existence of negative temperature in bounded systems
 \cite{Rapp10,Braun13,Ne_Temp_thr1,Ne_Temp_thr2,Ne_Temp_thr3,Ne_Temp_thr4,Ne_Temp_exp1},
 debates have been seen \cite{Dunkel13,Dunkel14,Julian}.
 To make the situation clarified, a direct consideration of the dynamics at the microscopic level should be unavoidable.

 On the other hand,  although at the macroscopic scale temperature can be detected in
 a reliable way by a thermometer,
 this strategy faces an obstacle, when applied to a small system which is coupled to a small probe as a micro thermometer.
 In fact, here, due to the smallness of the studied system, the system-probe interaction
 may impose nonnegligible influences in the measurement result, particularly, due to factors such as
 the form, location, and strength of the system-probe coupling,
 as well as the Hamiltonian and initial state of the probe.
 A reliable temperature detection can be accomplished, when these influences can be suppressed.

 With these considerations,  in this paper, we propose a temperature-detection method,
 which is based on an analysis of the dynamical evolution of the system-probe composite
 and gives a result insensitive to all the factors discussed above.
  A close relationship between the statistical mechanics and chaos has been perceived for a long time
 \cite{RMT_Casati,LL-book,Bannur97,Berdich}.
 Hence, we consider possible temperature detection for small quantum chaotic systems.
 With a two-level system employed as a probe,
 we'll show that the above-discussed insensitivity can indeed be achieved in certain situation.
 Interestingly,  it is found that the Boltzmann temperature can appear in a natural way from
 the dynamical evolution of the composite system.

 According to recent progresses achieved in the foundation
 of quantum statistical mechanics,
 the so-called typical states within appropriately-large energy shells
 have many properties similar to equilibrium states
 \cite{PSW06,Gold06,Haake,Izrailev12,Reimann08,LPSW09,
 Eisert12_notweakcoupling,pre12-stat,Eisert16_reivew}.
 For this reason, we consider a typical state of the studied system as its initial state,
 before it is coupled to the probe.
 It is found that, to accomplish the temperature detection, the system-probe coupling should be appropriately adjusted;
 specifically, the coupling should be able to induce chaotic motion of the total system,
 while, it should be still weak to ensure narrow eigenfunctions of the total system.
 The analytical results will be tested numerically in an Ising chain
 in a nonhomogeneous transverse field.

 The paper is organized as follows.
 In Sec.\ref{sect-setup}, we introduce the main setup.
 In Sec.\ref{sect-temp}, we discuss a reliable method of temperature detection
 for small quantum chaotic systems and derive an expression for the temperature thus determined.
 Then, we test the analytical predictions by numerical simulations in Sec.\ref{sect-numerical}.
 Finally, conclusions are given in Sec.\ref{sect-concl}.

 \section{The main setup}\label{sect-setup}

 We use $S$ to denote a considered quantum chaotic system
 and use $|\varphi_k\ra$ to denote eigenstates of its Hamiltonian $H_S$,
 $H_S |\varphi_k\ra = E_k |\varphi_k\ra$.
 As a quantum chaotic system, its spectrum has no degeneracy.
 Initially, the system $S$ lies in a (normalized) typical state
 within an energy shell $\Gamma_0$, centered at $E_S^{0}$ with a given
 width $\delta E$, namely, $\Gamma_0 = [E_S^{0}-\delta E/2,E_S^{0}+\delta E/2]$.
 Explicitly, the typical state is written as
\be
 |\Phi_0\rangle = \sum_{E_{k} \in \Gamma_0} D_{k} |\varphi_k\rangle,
\ee
 where $D_k$ are Gaussian random numbers with a same variance.
 We use $N_{\Gamma_0}$ to indicate the number of energy levels in the energy shell $\Gamma_0$.

 When a probe is coupled to the system $S$, the total Hamiltonian is written as
\begin{equation}\label{}
  H = H_p +\lambda H_I + H_S,
\end{equation}
 with a parameter $\lambda$ for adjusting the coupling strength.
 We use $|m\ra$ of $m=0,1$ to denote eigenstates of the probe Hamiltonian $H_p$
 with eigenvalues $e_m$, $H_p|m\ra = e_m|m\ra$.
 For brevity, we write unperturbed states  of the total system as
 $|\varphi_{k} m\ra$ with energies $E_{km} \equiv E_k + e_m$.
 Eigenstates of the total Hamiltonian $H$ are denoted by $|\psi_\alpha\ra$
 with energies $E_{\alpha}$, $H|\psi_{\alpha}\rangle=E_{\alpha}|\psi_{\alpha}\rangle$,
 and are expanded as
\begin{equation}\label{}
 |\psi_{\alpha}\ra  =  \sum_{k,m}C_{km}^{\alpha}|\varphi_{k} m\ra
\end{equation}
 in the unperturbed basis.
 The initial state of the total system is taken as $|\Psi_0\ra =  |\Phi_0\ra |m_0\ra$,
 undergoing a unitary evolution, $|\Psi(t)\ra = e^{-iHt}|\Psi_0\ra$.

 For it to be possible to use properties of the probe to detect properties of the system $S$ such as temperature,
 the motion of the probe should be sufficiently influenced by that of the system $S$.
 This requires that the probe-system coupling should not be very weak.
 Below, we assume that the probe is sufficiently coupled to the system,
 such that the total system is also a quantum chaotic system.
 (We'll revisit this point when discussing numerical results.)

 When the total system is a quantum chaotic system, its energy levels, as well as their spacings,
 have no degeneracy.
 It is known that, in this situation, the distance between the reduced density matrix (RDM) of the probe,
 $\rho(t) = {\rm Tr}_S(|\Psi(t)\ra \la \Psi(t)|)$, and its long-time average,
 denoted by $\ov\rho$, scales as $N_{\Gamma_0}^{- 1/2}$
 \cite{VoN,LPSW09,Reimann08,Reimann16,Reimann_relaxtime,Short11,Short12}.
 This implies that, at large $N_{\Gamma_0}$,
 if $\rho(t)$ has a steady state, it should be $\ov\rho$.

 To derive an expression for $\ov \rho$,
 we note that, when the RDM of the probe is measured experimentally,
 many realizations of the initial state of the system should be involved.
 Averaging over these initial states gives
 $\overline{D_{k_{0}}D_{l_{0}}}=\frac{1}{N_{\Gamma_{0}}}\delta_{k_{0}l_{0}}$.
 Then, taking average over a long-time period,  direct derivation shows that (cf.,e.g., Ref.\cite{LPSW09})
\begin{gather} \label{rhomm-steady}
 \ov\rho_{mm}
=  \frac{1}{N_{\Gamma_{0}}} \sum_{E_{k_0} \in \Gamma_0}  \sum_{k, \alpha}
 |C_{k_{0} m_0}^{\alpha}|^{2} |C_{km}^{\alpha}|^{2}.
\end{gather}
 Let us write $\ov\rho_{mm}$ as
\be\label{rho-Pkm}
 \ov\rho_{mm} =  \sum_{k}P_{m}^{m_0}(E_k),
\ee
 where
\begin{gather}\label{Pm-m0-Ek}
 P_{m}^{m_0}(E_k)= \frac{1}{N_{\Gamma_{0}}} \sum_{E_{k_0}\in \Gamma_0} P_{km}^{k_{0}m_{0}},
 \\
 P_{km}^{k_{0}m_{0}} \equiv \sum_{\alpha}|C_{k_{0}m_{0}}^{\alpha}|^{2}|C_{km}^{\alpha}|^{2}.
\end{gather}
 The quantity $P_{km}^{k_{0}m_{0}}$ has a simple interpretation, that is,
 it is the overlap of two local spectral density of states (LDOS).
 Specifically, defining a LDOS for an unperturbed state $|\varphi_km\ra$  as
 $\rho^{\rm L}_{km}(E) = \sum_\alpha |C_{km}^{\alpha}|^{2} \delta(E-E_\alpha)$ \cite{Flaum,WIC96},
 $P_{km}^{k_{0}m_{0}}$ is the overlap of $\rho^{\rm L}_{km}(E)$ and
 $\rho^{\rm L}_{k_0m_0}(E)$.
 Although the overlap $P_{km}^{k_{0}m_{0}}$ may show considerable fluctuations with variation
 of the system's energy $E_k$,
 the averaged overlap $P_{m}^{m_0}(E_k)$ should show a smoother feature
 for $N_{\Gamma_0}$ not small.

 We note that, for large $N_{\Gamma_0}$, off-diagonal elements of $\ov\rho$ can be neglected.
 In fact, applying a result given in Ref.\cite{pre14-ps}
 to the system-probe composite we study here with ${\rm Tr}_S(H_I) =0$,
 one finds that the steady state of the probe should have an approximately-diagonal form in the
 eigenbasis $\{ |m\ra\}$ at $\lambda$ not small,
 with off-diagonal elements scaling as $N_{\Gamma_0}^{- 1/2}$.
 In other words, the eigenbasis of the self-Hamiltonian of the probe is a preferred basis \cite{Zurek-ps,pra08-ps}.

 \section{Temperature detection}\label{sect-temp}

  For a probe as a two-level system, which has interacted with the measured system $S$
  and has reached a steady state $\ov\rho$,
  one can always get  a value of $\beta$ by fitting the steady state
  $\ov\rho$ to the canonical state $\frac 1Z e^{-\beta H_p}$.
  This value of $\beta$ reflects a property of the total system after the interaction.
  The point is  whether it is possible to determine certain value of $\beta$,
  which reflects a property of the initial state of the system $S$.
 For this to be possible, the finally determined value of $\beta$
 should be sensitive to neither the form, location, and strength of
 the probe-system coupling, nor the Hamiltonian and initial state of the probe.

 In this section, we show that the above-discussed goal can be achieved.
 That is, under appropriate conditions, a value of $\beta$
 can be obtained, which is insensitivity to the factors mentioned above.


\subsection{Properties of the function $P_{m}^{m_0}(E_k)$}

 In this subsection, we discuss properties of the function $P_{m}^{m_0}(E_k)$,
 which are useful in the study of the steady state $\ov\rho$ in Eq.(\ref{rho-Pkm}).
 As mentioned previously, the total system is assumed to be a quantum chaotic system,
 which implies that the eigenfunctions have sufficiently irregular components in the unperturbed basis.
 This chaotic feature requires that the coupling-strength $\lambda$ is not very small.
 Meanwhile, we require that $\lambda$ is not large, such that both eigenfunctions and LDOS are narrow
 with $w_L \ll \delta E$, where $w_L$ is the averaged width of LDOS, which is approximately equal to the
 averaged width of eigenfunctions for $\lambda $ not large.

 We find that, under the conditions discussed above,
 the function $P_{m}^{m_0}(E_k)$ has the following three properties.
 That is, (i) for a fixed value of $m_0$, this function with $m=0$ and with $m=1$
 have similar shapes, centered at $(E_{0}^S + e_{m_0}-e_m)$,
 (ii) it has a width approximately equal to $\delta E$,
 and (iii) it is approximately symmetric with respect to its center.

 To show the above-discussed properties, let us first consider the sum
\begin{gather}\label{Xm0}
 X_{m_0}(E_\alpha)\equiv \sum_{E_{k_0}\in \Gamma_0}|C_{k_{0}m_{0}}^{\alpha}|^{2},
\end{gather}
 as a function of the energy $E_\alpha$.
 Using this quantity, $P_{m}^{m_0}(E_k)$ in Eq.(6) can be written as
\begin{gather}\label{Pmm0-X}
 P_{m}^{m_0}(E_k)= \frac{1}{N_{\Gamma_{0}}} \sum_{\alpha}
 X_{m_0}(E_\alpha) |C_{km}^{\alpha}|^{2}.
\end{gather}
 The sum $X_{m_0}(E_\alpha)$ can be divided into a smoothly-varying part, denoted by
 $F_{m_0}(E_{\alpha})$, and a fluctuating part denoted by ${R_\alpha}$,
\begin{equation}\label{sum-C-FR}
 X_{m_0}(E_\alpha) = F_{m_0}(E_{\alpha}) +R_\alpha.
\end{equation}

 In the case of $\lambda =0$, there is a one-to-one correspondence between
 the set $\{|\psi_\alpha\}$ and the set $\{|\varphi_k m\ra \}$.
 To indicate this correspondence explicitly, we write the labels $k$ and $m$
 as ${k_{\alpha}}$ and $m_\alpha$.
 It is easy to verify that, at this $\lambda =0$,
\begin{gather}\label{}
  X_{m_0}(E_\alpha) =
 \left\{ \begin{array}{ll}
 1, & \text{if $E_{k_\alpha} \in \Gamma_0 \ \& \ m_\alpha = m_0$}
 \\ 0,  & \text{otherwise},
 \end{array} \right.
\end{gather}
 where $E_{k_\alpha} = E_\alpha - e_{m_\alpha}$.
 This implies that
\begin{gather}\label{eq-FEalpha}
  F_{m_0}(E_\alpha) =  \left\{ \begin{array}{ll}
 \frac{\rho_S(E_\alpha - e_{m_0})}{\sum_m\rho_S(E_\alpha-e_m)}, & \text{if $E_{k_\alpha} \in \Gamma_0$ }
 \\  0,  & \text{otherwise},
 \end{array} \right.
\end{gather}
 and
\begin{gather}  R_\alpha =  \left\{ \begin{array}{ll}
 1-\frac{\rho_S(E_\alpha - e_{m_0})}{\sum_m\rho_S(E_\alpha-e_m)}, & \text{if $E_{k_\alpha} \in \Gamma_0 \ \& \ m_\alpha = m_0$}
 \\  -\frac{\rho_S(E_\alpha - e_{m_0})}{\sum_m\rho_S(E_\alpha-e_m)},  & \text{if $E_{k_\alpha} \in \Gamma_0 \ \& \ m_\alpha \ne m_0$}
 \\ 0, & \text{otherwise}.
 \end{array} \right. \label{R-lambda=0}
\end{gather}
 As the probe is much smaller than the system, the function
 $F(E_\alpha)=\frac{\rho_S(E_\alpha - e_{m_0})}{\sum_m\rho_S(E_\alpha-e_m)}$
 usually changes quite slowly in the energy regions of interest.
 Then, one has
\begin{gather}\label{}
  F_{m_0}(E_\alpha)  \simeq  \left\{ \begin{array}{ll}
 c, & \text{if $E_{k_\alpha} \in \Gamma_0$ },
 \\  0,  & \text{otherwise},
 \end{array} \right.
\end{gather}
 where $c$ is some constant.
 Thus, the function $F_{m_0}(E_\alpha)$ has approximately a rectangular shape,
 centered at $E_S^0 + e_{m_0}$ with a width $\delta E$.
 In the case that the probe is a single qubit, whose energy scale is much smaller
 than that of the system $S$, one has
 $\rho_S (E_\alpha-e_1)\approx \rho_S (E_\alpha -e_0)$ and $c\approx \frac{1}{2}$.


 For small $\lambda$, the shape of the function $F_{m_0}(E_{\alpha})$
 should have only small deviation from that of $\lambda=0$ discussed above.
 Specifically, it should have the following properties:
 (i) being approximately symmetric with respect to
 a center $(E_S^0 + e_{m_0})$, (ii) having a width close to $\delta E$,
 (iii) varying slowly in the central region of its main body,
 and (iv) dropping fast at the edges to quite small values.
 Moreover, the main body of $R_\alpha$ should approximately lie in the same region
 as that of $F_{m_0}(E_{\alpha})$ discussed above.

 Since the total system is a quantum chaotic system, which has irregular components
 in the main bodies of its eigenfunctions,
 the fluctuating part $R_\alpha$ should fluctuate irregularly.
 Its contribution to the r.h.s. of Eq.(\ref{Pmm0-X}) scales as $1/N_{\Gamma_0}^{1/2}$.
 Hence, for large $N_{\Gamma_0}$, one gets
\be \label{Pkm-1}
P_{m}^{m_0}(E_k) \simeq \frac{1}{N_{\Gamma_0}}\sum_{\alpha}F_{m_0}(E_{\alpha})|C_{km}^{\alpha}|^{2}.
\ee

\begin{figure}
\includegraphics[width=1\columnwidth]{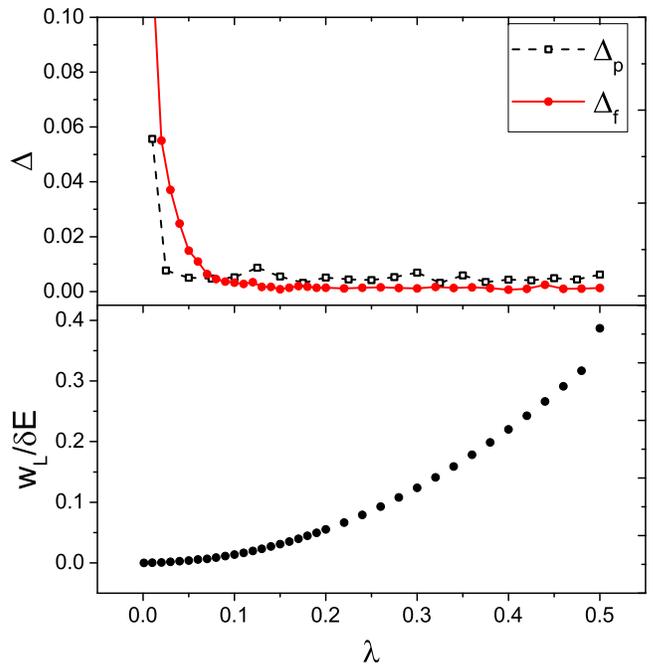}
\caption{(Color online) Upper panel:
  ``Distances'' to quantum chaos for the total system versus the coupling strength $\lambda$ for $N=14$.
  The distance $\Delta_p$ (see the text) ({empty squares connected by dashed line})  indicates a measure given by
  the statistics of spectrum and $\Delta_f$ ({solid circles connected by solid line}) 
  is for the statistics of eigenfunctions.
  Lower panel: the ratio  $w_L / \delta E$ versus $\lambda$.
  }
\label{diswl}
\end{figure}
 In the case of $w_L \ll \delta E$, for most of the LDOS $\rho^L_{km}(E_\alpha)$,
 their main bodies should lie within the slowly-varying region of
 the function $F_{m_0}(E_\alpha)$.
 For these LDOS, when computing the r.h.s. of Eq.(\ref{Pkm-1}),
 the term $F_{m_0}(E_\alpha)$ can be approximately taken as a constant.
 Then, noting that $\sum_\alpha |C_{km}^{\alpha}|^{2} =1$ and the fact that
 a narrow LDOS $\rho^L_{km}(E_\alpha)$ is approximately centered at $E_\alpha = E_{km}$,
 from Eq.(\ref{Pkm-1}) one finds that
\begin{gather}\label{Pkm0}
P_{m}^{m_0}(E_k)\simeq \frac{1}{N_{\Gamma_0}}F_{m_{0}}(E_\alpha)|_{E_\alpha = E_{km}}
\end{gather}
 for most of the energies $E_k$.
 Thus, for most of the LDOS $\rho^{\rm L}_{km}(E)$, the function $P_{m}^{m_0}(E_k)$ has the three
 properties stated above.

\subsection{Insensitivity to the coupling}

 In this section, making use of results given in the previous section,
 we show that a value of $\beta$ can be determined, which is insensitive to the coupling term
 under the conditions given previously.

 Substituting the expression of $P^{m_0}_m (E_k)$ in Eq.(\ref{Pkm0}) into Eq.(\ref{rho-Pkm}),
 one finds that, within an error with an upper bound of the order of $({w_L}/{\delta E})$,
\be
\ov\rho_{mm} \simeq   \frac{1}{N_{\Gamma_0}}\sum_{k}P_{m}^{m_0}(E_k)\simeq  \frac{1}{N_{\Gamma_0}}\sum_k F_{m_0}(E_{km}).
\ee
 If $\rho_{S}(E)$ can be approximated by a linear
 function within the energy shells centered at $(E_S^0+e_{m_0} - e_m)$ with a width $\delta E$,
 then, one gets
\bey\label{rho-Frho}
 \ov\rho_{mm}
& \simeq & G_{\lambda m_0}\ \rho_{{S}}(E_S^0+e_{m_0} - e_m),
\eey
 where
\begin{gather}\label{}
 G_{\lambda m_0} =  \frac{1}{N_{\Gamma_0}}F_{m_{0}}(E_{0}^S+e_{m_0}) \delta E,
\end{gather}
 being a quantity independent of the label $m$.
 The error for the approximation in Eq.(\ref{rho-Frho}) scales as
 $1/N_{\Gamma_0}^{1/2}$ and also as $(w_L/\delta E)$.
 Equation (\ref{rho-Frho}) predicts that
\begin{gather}\label{beta-m0}
 \beta
 \simeq\frac1{\Delta_e} \ln\frac{\rho_{{S}}(E_S^0+e_{m_0} - e_0)}{\rho_{{S}}(E_S^0
 +e_{m_0} - e_1)},
\end{gather}
 where $\Delta_e =  e_{1} - e_0$.
 It is clearly that the r.h.s. of Eq.(\ref{beta-m0}) is independent of
 the coupling term $\lambda H_I$.

 In the case that the eigenfunctions of the total system
 have on average a Lorentz shape \cite{EF-Flb},
 one can derive an explicit expression for the function $ P_{m}^{m_0}(E_k)$ (see {Appendix A}),
\bey \label{Pmm}
 P_{m}^{m_0}(E_k) \approx
 \frac{\left(\theta_+  - \theta_- \right) \rho_S(E_S^{0})  }{\pi\rho_T(E_S^{0}+e_{m_0})},
 \label{FE}
\eey
 where $\theta_\pm = \arctan \frac{2x_0 \pm \delta E}{2w_{L}}$, with
 $x_0 = E_k+e_m-E_S^{0}-e_{m_0}$, and $\rho_T$ is the
 density of states of the total system.
 It is not difficult to verify that the r.h.s. of Eq.(\ref{Pmm}) has the three properties
 discussed above for $P_{m}^{m_0}(E_k)$.

 \begin{figure}
\includegraphics[width=\columnwidth]{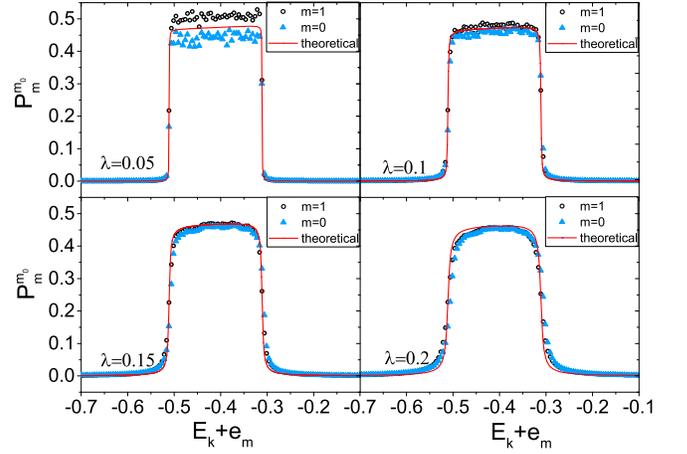}
\caption{(Color online)  Shapes of $P_{m}^{m_0}(E_k)$    for $m=0$ (empty circles) and $m=1$ (triangles),
   with $\delta E =0.2$ and $\Delta_e =0.6$,
 plotted as a function of $E_{km}$ for clearness in comparison.
   The solid curves represent the analytical prediction in Eq.(\ref{FE}).
   }
\label{EFchaos}
\end{figure}

\subsection{Insensitivity to the probe}

 The value of $\beta$ given in Eq.(\ref{beta-m0}) depends on both the initial state
 and the Hamiltonian of the probe.
 In this section, we determined a value of $\beta$, which is independent of the these two factors.

 With the dependence on $m_0$ written explicitly, $\beta_{m_0}$ in Eq.(\ref{beta-m0})
 has the following explicit expressions,
\begin{gather}\label{beta-01}
 \beta_{0} \simeq\frac1{\Delta_e} \ln\frac{\rho_{{S}}(E_S^0)}{\rho_{{S}}(E_S^0-\Delta_e)},
 \quad \beta_{1} \simeq\frac1{\Delta_e} \ln\frac{\rho_{{S}}(E_S^0+\Delta_e)}{\rho_{{S}}(E_S^0)}.
\end{gather}
 It is seen that the average $\ov\beta = \frac 12 (\beta_{0} + \beta_{1})$
 satisfies the following relation,
\begin{gather}\label{ov-beta}
 \ov\beta  \simeq \beta_{\rm sm},
\end{gather}
 where $ \beta_{\rm sm}$  is a Boltzmann temperature,
 given in statistical mechanics for macroscopic systems from Boltzmann's entropy \cite{Landau},
\begin{gather}\label{beta-sm}
 \beta_{\rm sm} =
 \left. \frac{\partial\ln\rho_S(E)}{\partial E} \right|_{E=E_S^0},
\end{gather}
 which is clearly independent of the probe.

 Furthermore, Eq.(\ref{ov-beta}) can be obtained under a more generic initial condition
 of the probe, namely, for $|\psi_0\rangle =\sum_m c_m|m\rangle$
 with a random relative phase between $c_0$ and $c_1$.
 In fact, in this case, within the second-order expansion of $\ln \rho_S$ with respect to $\Delta_e$,
 one can show that (see {Appendix B})
\begin{equation}
\label{Temp_arbi} \beta \simeq \sum_{m_0} |c_{m_0}|^2\beta_{m_0}.
\end{equation}
 Then, taking average over all possible values of $|c_{m_0}|^2$, one gets the same
 averaged value of $\beta$ as in Eq.(\ref{ov-beta}).

 To summarize, when the following conditions are satisfied,
 a temperature $\ov\beta \simeq \beta_{\rm sm}$
 can be assigned to a quantum chaotic system $S$,
 which can be detected by a probe qubit.
 That is,
 (i) $N_{\Gamma_0}$ for the initial state of $S$ is sufficiently large;
 (ii) the total system is a quantum chaotic system,
 whose eigenfunctions have sufficiently irregular coefficients in the unperturbed basis;
 (iii) $w_L \ll \delta E$; and
 (iv) $\delta E$ is sufficiently small for linear approximation of $\rho_S(E)$ within related energy shells.
\begin{figure}
\includegraphics[width=1\columnwidth]{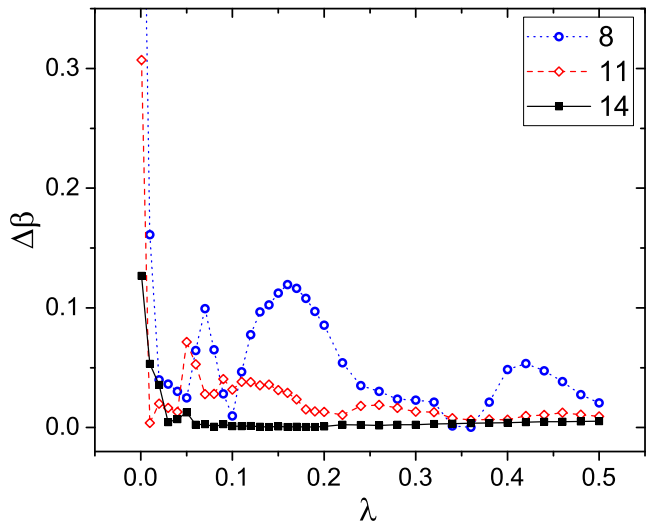}
\caption{(Color online)
  The difference $\Delta \beta = |\ov\beta -\beta_{\rm sm}|$
  versus $\lambda$. The value of $E_S^0$ for the initial state corresponds to $\beta_{\rm sm} =0.3$.
  }\label{NV}
\label{NV}
\end{figure}

 \section{Numerical tests}\label{sect-numerical}

 In this section, we test the results given above,
 by numerical simulations performed in an Ising chain composed of $N$ $\frac 12$-spins in a
 nonhomogeneous transverse field.
 The Hamiltonian of the system is written as
\be
 H_{S}=   \mu_{x} \sum_{i=1}^N\sigma_{x}^{i}
 +\mu_{1}\sigma_{z}^{1}+ \mu_{4}\sigma_{z}^{4}
 + \mu_{z}\sum_{i=1}^{N-1} \sigma_{z}^{i}\sigma_{z}^{i+1},
\ee
 where $\sigma_{x,z}$ indicate Pauli matrices.
 The probe, with a Hamiltonian $H_p = \omega_{p}\sigma^p_{x}$, is coupled to
 the $i$-th spin of the Ising chain, with an interaction Hamiltonian,
\be
 \lambda H_I= \lambda \sigma^p_{z}\otimes\sigma_{z}^i.
\ee
 The energy shell for the initial state is chosen narrow, but containing
 a large number of levels.
 For $N=14$, $N_{\Gamma_0}$ is about $500$.

 The parameters $\mu_x, \mu_z, \mu_1$, and $\mu_2$ are adjusted,
 such that the system $S$ is in a quantum chaotic regime, in which the nearest-level-spacing
 distribution $P(s)$ is close to the Wigner distribution
 $P_W(s)=\frac{\pi}{2}s\exp(-\frac{\pi}{4}s^2)$,
 the latter of which is almost identical to
 the prediction of the random matrix theory (RMT) \cite{RMT_Casati,RMT_Haake}.
 In order to determine the quantum chaotic regime of the coupling strength $\lambda$,
 we have studied the distance between $P(s)$ and $P_W(s)$, measured by
 $\Delta_p=\int |I(s)-I_{W}(s)|ds$.
 Here, $I(s)$ indicates the cumulative distribution of $P(s)$, $I(s) = \int_0^s P(s')ds'$,
 and $I_W(s)$ is the cumulative Wigner distribution, $I_W(s) = \int_0^s P_W(s')ds'$.
 As seen in the upper panel of Fig.\ref{diswl}, $\Delta_p$ drops quite fast, reaching a quite small
 value at $\lambda \approx 0.025$.

 As seen in the analytical derivation of temperature given in the previous section,
 the property, which has been really used, is certain irregular behavior of the eigenfunctions.
 Such a property of eigenfunctions is not necessarily guaranteed by properties of the spectrum.
 Hence, a direct study of statistical properties of the eigenfunctions is needed.
 Numerical simulations in several models, including the Ising chain studied here, show that
 the following quantity $\Delta_f$ is useful for this purpose \cite{chaosdis},
 $\Delta_f=\int |f(x)-f_{RMT}(x)|dx$.
 Here, $f(x)$ indicates the distribution of rescaled components in main bodies of
 the eigenfunctions, with $x=C_{km}^{\alpha}/ \sqrt{\Pi_m(\varepsilon)}$,
 where $\Pi_m(\varepsilon) = \la |C_{km}^{\alpha}|^2 \ra$
 indicates the average shape of the eigenfunctions
 and $f_{RMT}(x)$ is a Gaussian distribution predicted by the RMT \cite{RMT_Haake}.
 As seen in the upper panel of Fig.\ref{diswl}, $\Delta_f$ reaches its
 lowest-value region at $\lambda \approx 0.1$.
 Thus,  for $\lambda \gtrsim 0.1$, the eigenfunctions should have
 the needed irregular behaviors.

 The lower panel of Fig.\ref{diswl} shows that $w_L$
 reaches $10\%$ of $\delta E$ at $\lambda \approx 0.25$.
 Thus, the averaged overlap $P_{m}^{m_0}(E_k)$ is expected to
 have the three properties stated previously  for $\lambda$
 above $0.1$ and somewhat below $0.25$.
 Indeed, we found that $P_{m}^{m_0}(E_k)$ are close to the prediction in Eq.(\ref{FE})
 and possess the three properties in this intermediate regime of $\lambda$ (as illustrated in Fig.\ref{EFchaos}).
 Consistently,  $\ov\beta =\frac{\beta_1 +\beta_0}{2}$ has been found quite close to
 $\beta_{\rm sm}$ in this regime of $\lambda$ for $N=14$ (Fig.\ref{NV}).

 Fig.\ref{NV} shows that, decreasing the value of $N$ and, thus, decreasing the number
 $N_{\Gamma_0}$ for energy levels in the initial energy shell,
 the fluctuation of $\ov \beta$ becomes stronger.
 In fact, for $N=8$, the fluctuations are quite strong, such that no reliable temperature detection
 can be done by the probe.
 Furthermore, at quite small $\lambda$, even for
 large $N$ ($N=14$), the fluctuation of $\ov\beta$ is also quite large,
 such that there is no reliable temperature detection.
 In fact, in this case,  the two systems are not sufficiently coupled, as a result,
 one can not get the temperature of the system $S$ from properties of the probe.
 (This point is obvious in the extreme case of zero coupling.)

\begin{figure}
\includegraphics[width=\columnwidth]{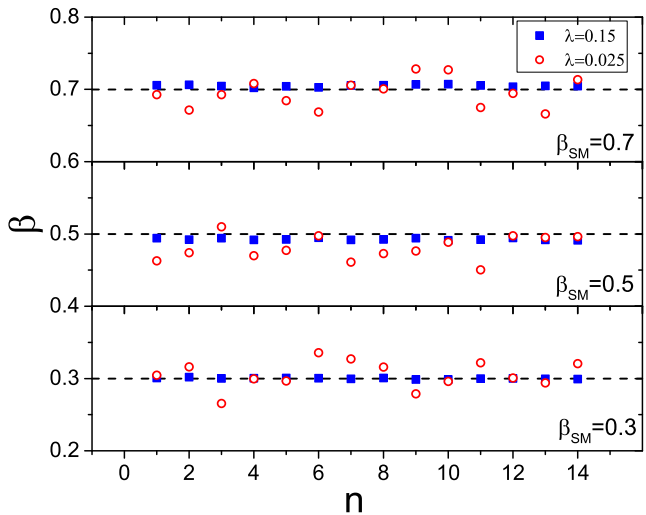}
\caption{Values of $\ov\beta$, when the probe is coupled to the $n$-th spin of the chain,
  for $\lambda =0.15$ (full squares) and for $\lambda=0.025$ (empty circles).
  }
\label{position}
\end{figure}

 We have also tested the insensitivity of the measured value $\ov \beta$
 to the location of coupling,
 for $\lambda$ lying in the intermediate regime discussed above, as illustrated in Fig.\ref{position}
 for $\lambda =0.15$.
 On the other hand, the figure shows that, for quite small values of $\lambda$, say, for $\lambda =0.025$,
 consistent with the results shown in Fig.\ref{NV},
 the value of $\ov \beta$ is sensitive to the location of coupling.
 Furthermore, we have studied dependence of the difference $|\beta_{1}-\beta_{0}|$ on the spin number $N$.
 The density of states $\rho_{{S}}$  has approximately
 a Gaussian shape \cite{DOSIsing}, $\rho_{{S}}(E) \approx A\exp(-\alpha E^{2})$.
 This gives $|\beta_{1}-\beta_{0}|\simeq2\alpha\Delta_e$.
 Numerically we found that $\alpha\propto \frac{1}{N+c}$ with
 $c\sim {\cal{O}}(1)$, hence, $|\beta_{1}-\beta_{0}|\propto \frac{\Delta_e}{N+c}$,
 approaching zero in the limit $ {N\rightarrow \infty}$.

 \section{Conclusions}\label{sect-concl}

 In this paper, it is shown that a probe qubit, which is appropriately coupled to a  small quantum chaotic system,
 can play the role of a microscopic thermometer.
 The obtained temperature is determined by the derivative of the logarithm of the density of states of the
 studied system, in the same manner as a Boltzmann temperature for macroscopic systems.
 The extent to which a temperature can be assigned to the system has also been studied.
 Finally, we note that the proposed method
 should be feasible for experimental study of temperature under nowadays technology.

\section{Acknowledgements}
 The authors are grateful to J. Gong, G. Casati, and G. Benenti for valuable discussions and suggestions.
 This work was partially supported by the Natural Science Foundation of China under Grant
 Nos.~11275179 and 11535011,
 and the National Key Basic Research Program of China under Grant
 No.~2013CB921800.

\begin{appendix}

\section{Derivation of Eq.(\ref{Pmm}) }
 In this appendix, we derive Eq.(\ref{Pmm}),
 when eigenfunctions of the total system have on average a Lorentz shape with a width $w_L$.
 In this case, one has
\be \label{C-Lorentz}
\overline{|C_{km}^{\alpha}|^{2}}\approx\frac{w_{L}}{\pi\rho_{T}(E_\alpha)}\cdot\frac{1}
{(E_{\alpha}-(E_{k}+e_{m}))^{2}+w_{L}^{2}},
\ee
 where the average is taken over neighboring levels\cite{Izrailev2000}.

 Noting Eqs.(\ref{Xm0}) and (\ref{sum-C-FR}), the
 smoothly-varying part $F_{m_{0}}(E_{\alpha})$ can be written as
\begin{gather}\label{Fm-ovC}
 F_{m_{0}}(E_{\alpha}) = \sum_{E_{k_{0}}\in\Gamma_{0}}\ov{|C_{k_{0}m_{0}}^{\alpha}|^{2}}.
\end{gather}
 When $N_{\Gamma_0}$ is large, the summation in Eq.(\ref{Fm-ovC}) can be approximated
 by an integration over the energy of the system $S$, with $\int dE \rho_S(E)$.
 Substituting Eq.(\ref{C-Lorentz}) into the obtained integration, one gets
\begin{gather}
F_{m_{0}}(E_{\alpha})
 \simeq \frac{\rho_{S}(E_S^0)w_{E}}{\rho_{T}(E_\alpha)\pi}
\int_{E_{\alpha}-E_S^0-e_{m_{0}}-\frac{\delta E}{2}}^{E_{\alpha}-E_S^0-e_{m_{0}}
+\frac{\delta E}{2}}\frac{1}{x^{2}+w_{E}^{2}}dx,
\end{gather}
 where $x = E_{\alpha}-(E+e_{m_0})$.
 Then, noting Eq.({\ref{Pkm0}}) and the fact that $E_{km} = E_k +e_m$,
 straightforward derivation shows that
\begin{gather}\notag
 P_{m}^{m_0}(E_k) 
 \simeq \frac{\rho_S(E_S^0)}{\pi\rho_T(E_T^0)}\left(\arctan\frac{2x_0+\delta E}{2w_{E}}
 \right.
 \\ \left.  -\arctan\frac{2x_0-\delta E}{2w_{E}} \right), \label{FE-app}
\end{gather}
 where $x_0=E_k+e_m-E_S^0-e_{m_0}$.

\section{Derivation of Eq.(\ref{Temp_arbi}) for a generic initial state of the probe}

 In this appendix, we show that Eq.(\ref{Temp_arbi}) holds, within the second-order
 approximation with respect to $\Delta_e =e_1-e_0$, under a generic initial condition
 of the probe, $|\psi_0\rangle =\sum_m c_m|m\rangle$
 with a random relative phase between $c_0$ and $c_1$.
 Below, for brevity, in this appendix we omit the overline of $\ov\rho$.

 Taking average over the initial states $|\psi_0\ra$, due to the random relative phase between $c_0$ and $c_1$,  one gets
\begin{gather}\label{rho-any}
\rho_{mm}
 = \sum_{m_0} |c_{m_0}|^2 \rho_{mm}^{(m_0)},
\end{gather}
 where $\rho_{mm}^{(m_0)}$ indicates the r.h.s.~of Eq.(\ref{rhomm-steady}),
 with the dependence on $m_0$ written explicitly.
 As discussed in the main text, the averaged RDM has an approximately-diagonal form
 in the eigenbasis of the self-Hamiltonian $H_p$.
 In this basis, the parameter $\beta$ in the canonical state
 $\frac{1}{Z}\exp(-\beta H_p)$ can written as
\be
\beta = -\frac{1}{\Delta_e}\ln\frac{\rho_{11}}{\rho_{00}}.
\ee
 Substituting Eq.(\ref{rho-any}) into the above expression of $\beta$,
 one gets
\be\label{beta-any}
\beta = -\frac{1}{\Delta_e}\ln\frac{|c_{1}|^2\rho_{11}^{(1)}+|c_{0}|^2\rho_{11}^{(0)}}
{|c_{1}|^{2}\rho_{00}^{(1)}+|c_{0}|^{2}\rho_{00}^{(0)}}.
\ee

 Making use of the expression of $\rho_{mm}^{(m_{0})}$ in Eq.(\ref{rho-Frho}), it is not difficult to find that
\be
\rho_{mm}^{(m_{0})}\simeq\frac{\rho_{S}(E_S^0+e_{m_{0}}-e_{m})}
{\sum_{m'}\rho_{S}(E_S^0+e_{m_{0}}-e_{m'})}.
\ee
 For example, for $m=m_0 =1$, one has
\be\label{rho-1}
\rho_{11}^{(1)}\simeq\frac{\rho_{S}(E_S^0)}{\rho_{S}(E_S^0)+\rho_{S}(E_S^0+\Delta_e)}.
\ee

 Expanding $\ln\rho_S(E_S^0+\Delta_e)$
 in the Taylor's expansion and keeping the second-order term, one finds that
\begin{gather} 
\rho_{S}(E_S^0+\Delta_e) 
  \simeq \rho_{S}(E_S^0)\exp(\beta_{\rm sm}\Delta_e +\beta_{\rm sm}'\Delta_e^2/2), \label{Dos}
\end{gather}
 where $\beta_{\rm sm}$ is defined in Eq.(\ref{beta-sm}),
 $\beta_{\rm sm}=\frac{\partial\ln\rho_{S}}{\partial E}|_{E=E_S^0}$.
 Substituting Eq.(\ref{Dos}) into Eq.(\ref{rho-1}), one gets
\be\label{rho11}
\rho_{11}^{(1)}\simeq\frac{1}{1+\exp(\beta_{\rm sm}\Delta_e +\beta_{\rm sm}'\Delta_e^2/2)}.
\ee
 Similarly, one can compute other elements $\rho_{mm}^{(m_0)}$.

 To simplify the notation, we introduce two quantities $\chi_{+}$ and $\chi_{-}$,
\begin{gather}\label{}
\chi_{+}=\exp(\beta_{\rm sm}\Delta_e+\beta'_{sm}\Delta_e^{2}/2),
\\ \chi_{-}=\exp(\beta_{\rm sm}\Delta_e-\beta'_{sm}\Delta_e^{2}/2).
\end{gather}
 It is not difficult to find that
\begin{gather}
\rho_{11}^{(1)} \simeq \frac{1}{1+\chi_{+}}, \ \rho_{00}^{(1)} \simeq \frac{\chi_{+}}{1+\chi_{+}},  \nonumber \\
\rho_{11}^{(0)} \simeq \frac{1}{1+\chi_{-}}, \ \rho_{00}^{(0)} \simeq \frac{\chi_{-}}{1+\chi_{-}}.
\end{gather}
 Substituting these expressions into Eq.(\ref{beta-any}),
 after simple algebra, we get
\begin{gather}
\beta \simeq -\frac{1}{\Delta_e}\ln\frac{|c_{1}|^{2}(1+\chi_{-})+|c_{0}|^{2}(1+\chi_{+})}
{|c_{1}|^{2}\chi_{+}(1+\chi_{-})+|c_{0}|^{2}\chi_{-}(1+\chi_{+})} \nonumber \\
 \simeq -\frac{1}{\Delta_e}\ln\frac{1+|c_{1}|^{2}\chi_{-}+|c_{0}|^{2}\chi_{+}}
{|c_{1}|^{2}\chi_{+}+|c_{0}|^{2}\chi_{-}+\exp(2\beta_{\rm sm}\Delta_e)}. \label{beta-app1}
\end{gather}

 When $(\beta_{\rm sm}'\Delta_e^{2})$ is small, one can write
\be
\exp(\beta_{\rm sm}'\Delta_e^{2}/2)\simeq 1+\beta_{\rm sm}'\Delta_e^{2}/2.
\ee
 Using this approximation, Eq.(\ref{beta-app1}) can be further written as
\begin{gather}
\beta \simeq \beta_{\rm sm} \hspace{6.5cm} \notag
 \\ -\frac{1}{\Delta_e}\ln\frac{1+\exp(\beta_{\rm sm}\Delta_e)
 [1-(|c_{1}|^{2}-|c_{0}|^{2})\beta'_{sm}\Delta_e^{2}/2]}{1+(|c_{1}|^{2}-|c_{0}|^{2})
 \beta'_{sm}\Delta_e^{2}/2+\exp(\beta_{\rm sm}\Delta_e)}.
\end{gather}
 Then, using the approximation that
\be 1\pm(|c_{1}|^{2}-|c_{0}|^{2})\beta'_{sm}\Delta_e^{2}/2
\simeq \exp(\pm(|c_{1}|^{2}-|c_{0}|^{2})\beta'_{sm}\Delta_e^{2}/2),
\ee
 straightforward derivation gives
\begin{gather} \notag
\beta \simeq \beta_{\rm sm} \hspace{7cm}
 \\ -\frac{1}{\Delta_e}\ln\frac{1+\exp(\beta_{\rm sm}\Delta_e)
\exp(-(|c_{1}|^{2}-|c_{0}|^{2})\beta'_{sm}\Delta_e^{2}/2)}{\exp((|c_{1}|^{2}-|c_{0}|^{2})
\beta'_{sm}\Delta_e^{2}/2)+\exp(\beta_{\rm sm}\Delta_e)} \nonumber \\
 = \beta_{\rm sm}+(|c_{1}|^{2}-|c_{0}|^{2})\beta'_{sm}\Delta_e/2 .
\end{gather}
 Finally, noting that $|c_1|^2+|c_0|^2=1$ and using the expressions of $\beta_0$ and $\beta_1$ in
 Eq.(\ref{beta-01}), one gets
\begin{gather}\label{}
 \beta \simeq |c_1|^2(\beta_{\rm sm}+\beta'_{sm}\Delta_e/2) +|c_0|^2
 (\beta_{\rm sm}-\beta'_{sm}\Delta_e/2) \nonumber \\
 \simeq |c_1|^2\beta_1 + |c_0|^2\beta_0,
\end{gather}
 which gives Eq.(\ref{Temp_arbi}).

\end{appendix}


\begin{thebibliography}{1}
\bibitem{Heatengine_thr2} N. Linden, S. Popescu, and P. Skrzypczyk, Phys. Rev. Lett. \textbf{105}, 130401(2010).
\bibitem{Heatengine_thr3}O. Fialko and D.W. Hallwood, Phys. Rev. Lett. \textbf{108}, 085303 (2012).
\bibitem{Heatengine_exp1} M. Serra-Garcia, \textit{et al.}, Phys. Rev. Lett. \textbf{117}, 010602 (2016).
\bibitem{Heatengine_thr1}H. T. Quan, Yu-xi Liu, C. P. Sun, and Franco Nori, Phys. Rev. E \textbf{76}, 031105 (2007).
\bibitem{gogolin}J. Eisert, M. Friesdorf, and C. Gogolin, Nature Physics \textbf{11}, 124 (2015).

\bibitem{Hartmann04} M. Hartmann, G. Mahler, and O. Hess, Phys. Rev. Lett. \textbf{93},080402 (2004).
\bibitem{Eisert14} M. Kliesch, C. Gogolin, M.J. Kastoryano, A. Riera, and J. Eisert, Phys. Rev. X \textbf{4},031019 (2014).

\bibitem{Rossini16} A. De Pasquale, D. Rossini, R. Fazio, and V. Giovannetti,
 Nat. Commun. {\bf 7}, 12782 (2016).


\bibitem{Ne_Temp_thr1} N. F. Ramsey, Phys. Rev. \textbf{103}, 20 (1956).
\bibitem{Ne_Temp_thr2} M. J. Klein, Phys. Rev. \textbf{104}, 589 (1956).
\bibitem{Ne_Temp_thr3}  A. S. Oja and O. V. Lounasmaa, Rev. Mod. Phys. \textbf{69}, 1 (1997).
\bibitem{Rapp10}  A. Rapp, S. Mandt, and A. Rosch, Phys. Rev. Lett. \textbf{105}, 220405 (2010).
\bibitem{Ne_Temp_exp1} P. Medley, D. M. Weld, H. Miyake, D. E. Pritchard, and W. Ketterle, Phys. Rev. Lett. \textbf{106}, 195301 (2011).
\bibitem{Braun13}  S. Braun, \textit{et al.}, Science \textbf{339}(6115), 52-55, (2013).
\bibitem{Ne_Temp_thr4} S. Mandt, A. E. Feiguin, and S. R. Manmana, 	Phys. Rev. A \textbf{88}, 043643 (2013).

\bibitem{Landau} L.D. Landau and E.M. Lifshitz, {\it Statistical Physcics}  (Pergamon, London, 1958).
\bibitem{Dunkel14} S. Hilbert, P. H\"{a}nggi, and J. Dunkel, Phys. Rev. E \textbf{90}, 062116 (2014).


\bibitem{Izrailev-old}V. V. Flambaum and F. M. Izrailev, Phys. Rev. E \textbf{55}, R13 (1997).
\bibitem{Casati-th1}G. Casati, B.V. Chirikov, I. Guarneri, and F. M. Izrailev, Phys. Lett. A \textbf{247}, 140 (1998).
\bibitem{Casati-th2}C. Mej\'{i}a-Monasterio, T.Prosen, and G. Casati, Europhys. Lett., \textbf{72}, 520  (2005).
\bibitem{Genway12} S. Genway, A. F. Ho and D. K. K. Lee, Phys. Rev. A \textbf{86}, 023609 (2012).
\bibitem{ethbath} O. Fialko, Phys. Rev. E \textbf{92}, 022104, (2015).
\bibitem{Izrailev-new}F. Borgonovi, F. Mattiotti, and F. M. Izrailev, arXiv:1608.01469.

\bibitem{Dunkel13} J. Dunkel and S. Hilbert, Nature Physics \textbf{10}, 67 (2014).
\bibitem{Julian} J. Poulter, Phys. Rev. E \textbf{93}, 032149 (2016).

 \bibitem{LL-book} A.J.~Lichtenberg and M.A.~Liebermann,  {\it Regular and Chaotic Dynamics},  2nd ed.,
 (Springer-Verlag, New York, 1992).
\bibitem{RMT_Casati} {\it Quantum Choas: Between Order
and Disorder}, edited by G.~Casati and B.V.~Chirikov
(Cambridge University Press, Cambridge, 1994).
\bibitem{Berdich}  V. L. Berdichevsky and M. V. Alberti, Phys. Rev. A \textbf{44}, 4858 (1991).
\bibitem{Bannur97} V. M. Bannur, P. K. Kaw, and J. C. Parikh, Phys. Rev.E \textbf{55}, 2525 (1997).

  \bibitem{PSW06} S. Popescu, A.J. Short, and A. Winter, {Nature Physics} {\bf 2}, 754-758 (2006).
 \bibitem{Gold06} S. Goldstein, J.L. Lebowitz, R. Tumulka, and N. Zanghi, {Phys. Rev. Lett.}  {\bf 96}, 050403 (2006).
\bibitem{Eisert12_notweakcoupling} Arnau Riera, Christian Gogolin, and Jens Eisert, Phys. Rev. Lett. \textbf{108}, 080402 (2012).
\bibitem{pre12-stat} W.-g.~Wang, Phys.~Rev.~E {\bf 86}, 011115 (2012).
\bibitem{Haake}  A. Altland and F. Haake,  Phys. Rev. Lett. \textbf{108},  073601 (2012).
\bibitem{Izrailev12} L.F. Santos, F. Borgonovi, and F.M. Izrailev, Phys. Rev. Lett. \textbf{108}, 094102 (2012).
\bibitem{Eisert16_reivew} C. Gogolin and J. Eisert, Rep. Prog. Phys. \textbf{79}, 056001 (2016).
\bibitem{Reimann08} P. Reimann, Phys. Rev. Lett. \textbf{101}, 190403 (2008).
 \bibitem{LPSW09} N. Linden, S. Popescu, A.J. Short, and A. Winter, {Phys. Rev. E} {\bf 79}, 061103 (2009).

\bibitem{VoN} J. Von Neumann, Beweis des Ergodensatzes und des H-Theorems in der neuen Mechanik, Zeitschrift f\"{u}r Physik \textbf{57}: 30-70 (1929).

\bibitem{Short11} A. J. Short, New J. Phys. \textbf{13}, 053009 (2011).
\bibitem{Reimann_relaxtime}P. Reimann, Phys. Scr. \textbf{86}, 058512 (2012).
\bibitem{Short12} A. J. Short and T. C Farrelly, New J. Phys. \textbf{14}, 013063  (2012).
\bibitem{Reimann16} P. Reimann, Nat. Commun. \textbf{7}, 10821 (2016).

\bibitem{Flaum} V. V. Flambaum, A. A. Gribakina, G. F. Gribakin, and M. G.
Kozlov, Phys. Rev. A \textbf{50}, 267 (1994).
\bibitem{WIC96} W.-g. Wang, F. M. Izrailev, and G. Casati, Phys. Rev. E \textbf{57}, 323 (1998).

\bibitem{pre14-ps} L. He and W.-g. Wang, Phys.Rev.E {\bf 89}, 022125 (2014).

\bibitem{Zurek-ps}  J. P. Paz and W. H. Zurek, Phys. Rev. Lett. \textbf{82}, 5181 (1999).
\bibitem{pra08-ps} W.-g. Wang, J.Gong, G.Casati, and B.Li, Phys.~Rev.~A {\bf 77},
 012108 (2008).
 derivations of Eqs.(\ref{rho-Frho}) and  (\ref{Pmm}), and about validity of Eq.(\ref{ov-beta}) under random initial states of the probe.

\bibitem{EF-Flb} V. V. Flambaum and F. M. Izrailev, Phys.~Rev.~E {\bf 61}, 2539 (2000).

\bibitem{RMT_Haake}  { F. Haake, \it Quantum Signatures of Chaos}, 3rd ed.,
(Springer-Verlag, Berlin, 2010).

\bibitem{chaosdis} J. Wang, W.-g. Wang, Chaos, Solitons \& Fractals, \textbf{91}, 291 (2016).
\bibitem{DOSIsing} Y.Y. Atas, E. Bogomolny,arXiv:1402.6858.
\bibitem{Izrailev2000} V.V.~Flambaum and F.M.~Izrailev, Phys.~Rev.~E {\bf 61}, 2539 (2000).
\end{thebibliography}
\end{document}